\documentclass[runningheads]{llncs}

\usepackage{url}
\usepackage{amsmath}
\usepackage{graphicx}

\begin{document}

\title{The evolution of AI approaches for motor imagery EEG-based BCIs}

\author{Aurora Saibene\inst{1,2}\orcidID{0000-0002-4405-8234} \and
	Silvia Corchs\inst{2,3}\orcidID{0000-0002-1739-8110} \and
	Mirko Caglioni\inst{1} \and
	Francesca Gasparini\inst{1,2}\orcidID{0000-0002-6279-6660} 
}
\authorrunning{A. Saibene et al.}

\institute{University of Milano-Bicocca, Viale Sarca 336, 20126, Milano, Italy \\
	\email{aurora.saibene@unimib.it, m.caglioni2@campus.unimib.it, francesca.gasparini@unimib.it}
	\and
	NeuroMI, Milan Center for Neuroscience, Piazza dell’Ateneo Nuovo 1, 20126, Milano, Italy \\
	\and	
	University of Insubria, Via J. H. Dunant 3, 21100, Varese, Italy\\
	\email{silvia.corchs@uninsubria.it}
}

\maketitle

\begin{abstract}
	The Motor Imagery (MI) electroencephalography (EEG) based Brain Computer Interfaces (BCIs) allow the direct communication between humans and machines by exploiting the neural pathways connected to motor imagination. Therefore, these systems open the possibility of developing applications that could span from the medical field to the entertainment industry. In this context, Artificial Intelligence (AI) approaches become of fundamental importance especially when wanting to provide a correct and coherent feedback to BCI users. Moreover, publicly available datasets in the field of MI EEG-based BCIs have been widely exploited to test new techniques from the AI domain. In this work, AI approaches applied to datasets collected in different years and with different devices but with coherent experimental paradigms are investigated with the aim of providing a concise yet sufficiently comprehensive survey on the evolution and influence of AI techniques on MI EEG-based BCI data.  
	
\end{abstract}

\section{Introduction}\label{sec:intro}
Translating thoughts into commands understandable by external applications and devices is the basic principle ruling the development of Brain Computer Interfaces (BCIs) \cite{singh2021comprehensive}. 
The most appreciated method to collect neural signals is the electroencephalogram (EEG), having that it records data with non-invasive surface sensors called electrodes, it is sufficiently low-cost and with possible high temporal and spatial resolution \cite{craik2019deep}. Moreover, the EEG signals are characterized by rhythms, whose fluctuations may be exploited to detect specific brain states \cite{vaid2015eeg}. Among these brain conditions, the imagination of voluntary movements, called Motor Imagery (MI), may be observed over the primary sensorimotor cortex with amplitude variations of the $\mu$ and $\beta$ rhythms \cite{wriessnegger2018frequency} \cite{dai2019eeg}. These effects can be exploited to create MI EEG-based BCIs that can be used for a variety of applications spanning from rehabilitation procedures to the control of wheelchair movements \cite{singh2021comprehensive}, and in conjunction with virtual and augmented reality \cite{kohli2022review}.\\
Therefore, one of the main BCI life-cycle components is represented by feedback, which benefits from the evolution and improvement of Artificial Intelligence (AI) approaches \cite{cao2020review} in predicting the different brain states to be translated into system commands.\\
Consequently, \textit{how have the AI techniques evolved in and influenced the field of MI EEG-based BCIs, considering the great number of possibilities offered by these systems?}\\ 
This work aims at providing a brief overview and discussion on this topic by analysing the AI approaches that have been applied to  some representative datasets present in the domain literature. 

Therefore, in Section~\ref{sec:overview} the paper firstly provides an overall view of the timeline related to MI EEG-based BCIs and justifies the choice of specific datasets, that are described in Section~\ref{sec:datasets}. An overview of the AI techniques applied to these data is provided in Section~\ref{sec:ai} and observations on the presented AI approaches are discussed in Section~\ref{sec:discussion}. Finally, conclusions are drawn in Section~\ref{sec:conclusion}.

\section{Overview}\label{sec:overview}
\begin{figure}
	\centering
	\includegraphics[width=\linewidth]{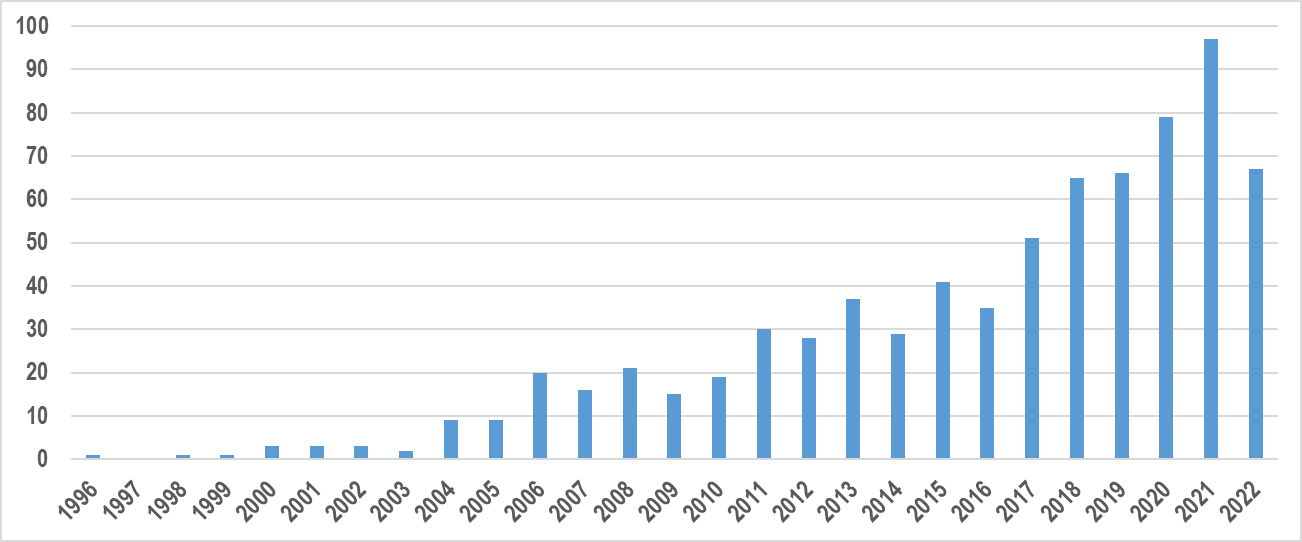}
	\caption{Number of papers per year obtained by querying Scopus on MI EEG-based BCI related keywords (search conducted during September 2022).}
	\label{fig:MI-EEG-based-BCI-query}
\end{figure}

The research on MI EEG-based BCIs has become particularly prolific in the last years.
In fact, typing a quick query on Scopus\footnote{\url{https://www.scopus.com/}} title, abstract and keywords of indexed works, i.e., \textit{TITLE-ABS-KEY((mi OR motor AND imagery OR  motor AND imagination)  AND  (eeg OR electroencephalographic) AND based AND (bci OR  brain AND computer AND interface))}, we obtain the graphic in Figure~\ref{fig:MI-EEG-based-BCI-query}. Notice that the search was conducted during September 2022.\\
Besides having a clear understanding of a constant increase of publications on these topics starting from 2017 and having for now its apex in 2021, Figure~\ref{fig:MI-EEG-based-BCI-query} provides a story of the early phases of the MI EEG-based BCIs, which set the foundation for later works.\\
An initial period (1996-2003) during which the BCI community began to discuss these topics was followed by a discrete boost (2004) of the research production probably due to the insightful work of \textit{Schalk et al.} \cite{schalk2004bci2000} and the outcome of the \textit{BCI Competition 2003} \cite{blankertz2004bci}.\\ 
Starting from the evident direction took by many laboratories concerning the development of systems capable of enabling a mean of communication and control for patients with severe motor disabilities, \textit{Schalk et al.} presented the BCI2000\footnote{\url{https://www.bci2000.org/mediawiki/index.php/Main_Page}}, which is an open-source platform that allows the management of BCI systems and that remains active and maintained to the present day. 
Moreover, not only some of the authors of \cite{schalk2004bci2000} were involved in the \textit{BCI Competition 2003}, but also the names of the researchers of the pioneer work of 1996 \cite{kalcher1996graz} obtained with the Scopus search, unsurprisingly appear. 
In fact, they contributed to dataset III titled \textit{Motor Imagery}, which presented data acquired on the main central cortical electrodes (C\{3,4,z\}) during the imagination of left or right hand movements, with the main aim of providing a continuous feedback to the BCI users. The winning strategy to the proposed problem was presented by \textit{Lemm et al.} \cite{lemm2004bci}, who tried to disclose motor intention by characterizing the EEG signal rhythmic activity through complex Morlet wavelet \cite{torrence1998practical} application and using a probabilistic model to predict left or right hand MI.

Afterwards, the \textit{BCI Competitions - Berlin Brain Computer-Interface}\footnote{\url{https://www.bbci.de/competition/}} datasets quickly became benchmarks on which test strategies to provide more efficient and reliable BCI systems.\\
By adding to the first query the string \textit{(bci AND competition AND  (2003 OR ii OR iii OR iv))}, representing the BCI Competition II (or 2003), III and IV, and limiting the results to year 2021 only, 27 papers are provided in output, of which 23 use the BCI Competition IV datasets (especially, 2a and 2b). Considering that Figure~\ref{fig:MI-EEG-based-BCI-query} reports 97 works published in 2021 and that the screening through Scopus is done only on the title, abstract, and keywords of the works indexed by this search engine, the number of publications using the \textit{BCI Competition IV dataset 2a} and \textit{2b} \cite{tangermann2012review} seems to be fairly high and justifies a deeper analysis concerning the AI techniques tested on them to better understand their evolution during a long time span.

However, these datasets present EEG recordings acquired on a restricted number of subjects, considering a restricted number of electrodes and experimental conditions (as detailed in Section~\ref{sec:datasets}). Wanting to have a general overview of the evolution of AI approaches in MI EEG-based BCIs and noticing its increased use as a benchmark in the last 10 years, the \textit{EEG Motor Movement/Imagery Dataset} \cite{schalk2004bci2000} \cite{goldberger2000physiobank} collected from a larger population, using a different montage, and considering diverse MI tasks, but using the BCI2000 system, is also considered. 

Moreover, a great attention has been given to wearable technologies, especially in the last few years, since the necessity of moving the use of BCIs from medical and laboratory environments to real-world scenarios is becoming more pressing due to a variety of needs like developing in-home rehabilitation tools \cite{daly2015brain}, exploiting customer-grade devices \cite{vasiljevic2020brain}, and providing continuous assistive technologies \cite{minguillon2017trends} that patients could easily use alone without having to buy expensive equipment.\\
Following these principles, \textit{Peterson et al.} \cite{peterson2020feasibility} have collected the \textit{MI-OpenBCI} dataset using wearable low-cost technologies to record MI EEG-based BCI data. Therefore, an overview of this dataset is provided to have a closer look on future developments of these new systems and the changing role of AI when facing them.


\section{Datasets}\label{sec:datasets}
Considering the brief overview presented in Section~\ref{sec:overview}, the datasets that will be at the center of this paper analysis are the \textit{BCI Competition IV dataset 2a} (2012) and \textit{2b} (2012) \cite{tangermann2012review}, the \textit{EEG Motor Movement/Imagery Dataset} (2009) \cite{schalk2004bci2000} \cite{goldberger2000physiobank}, and the \textit{MI-OpenBCI} (2020) one \cite{peterson2020feasibility}. In this section a concise description of their characteristics is reported for completeness.


\subsection{BCI Competition IV dataset 2a}
\begin{figure}
	\centering
	\includegraphics[width=0.7\linewidth]{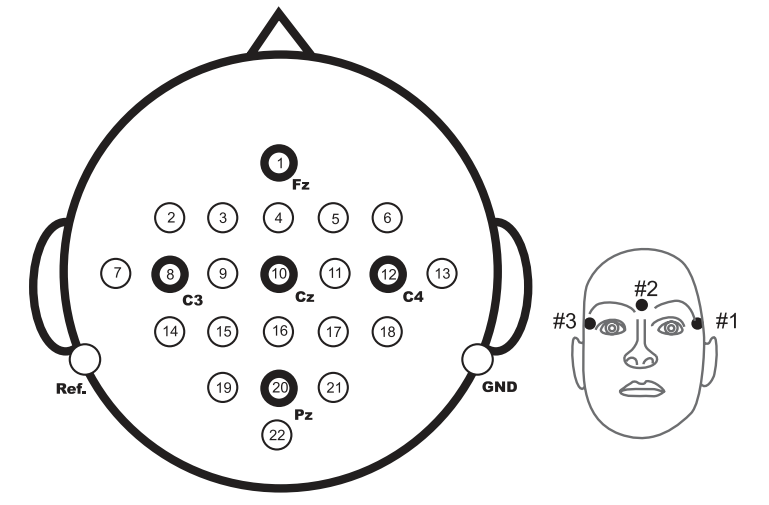}
	\caption{Electrode positioning used for the recording of BCI Competition IV dataset 2a. Notice that on the right are reported the positions of the electrooculogram channels. This setting is also used for BCI Competition IV dataset 2b, considering electrodes C\{3,4,z\} only. Original montage taken from \cite{tangermann2012review}.}
	\label{fig:montage-BCI-CompetitionIV-dataset2a}
\end{figure}

The \textit{BCI Competition IV dataset 2a} \cite{tangermann2012review} is provided under the name \textit{Continuous Multi-class Motor Imagery}. In fact, it has been collected from 9 subjects executing a cue-based BCI paradigm consisting of left/right hand, both feet and tongue MI. Each subject participated in different days to 2 experimental sessions containing 6 experimental runs of 48 trials each.\\ 
Figure~\ref{fig:montage-BCI-CompetitionIV-dataset2a} depicts the montage consisting of 22 Ag/AgCl electrodes for scalp recording, reference and ground electrodes placed on the left and right mastoids, and 3 monopolar electrooculogram channels (positioned to provide reference for artifact removal).\\
The signals have been acquired with 250Hz sampling rate and the dataset authors provided them bandpass (0.5-100Hz) and notch (50Hz) filtered. The  noisy trials were removed by manual screening of experts.

\subsection{BCI Competition IV dataset 2b}
The \textit{BCI Competition IV dataset 2b} \cite{tangermann2012review} presents the following short description: \textit{Session-to-Session Transfer of a Motor Imagery BCI under Presence of Eye Artifacts}.
In fact, the motivation driving its presence into the BCI competition was to provide a correct EEG signal classification despite having data affected by ocular noise.\\
The dataset has been collected from 9 right-handed healthy subjects using only the central cortical electrodes (C\{3,4,z\}), the reference and ground channel and the 3 monopolar electrooculogram channels (depicted in Figure~\ref{fig:montage-BCI-CompetitionIV-dataset2a}), during a cue-based BCI paradigm of left/right hand MI \cite{leeb2007brain}. The experiment was designed to have 2 separate sessions consisting of 6 runs of 10 trials each without user feedback and 3 separate sessions consisting of 4 runs of 40 trials each with online feedback.\\
The EEG signal was sampled at 250Hz and bandpass (0.5-100Hz) and notch (50Hz) filtered.

\subsection{EEG Motor Movement/Imagery Dataset}
The \textit{EEG Motor Movement/Imagery Dataset}\footnote{Dataset description and recordings available at \url{https://physionet.org/content/eegmmidb/1.0.0/}.} \cite{schalk2004bci2000} \cite{goldberger2000physiobank} has been collected by using the BCI2000 system introduced in Section~\ref{sec:overview} and considering an electrode montage of 64 channels respecting the 10-10 International System (excluding electrodes Nz, F\{9,10\}, FT\{9,10\}, A\{1,2\}, TP\{9,10\}, P\{9,10\}).\\
109 subjects were asked to perform a cue-based experiment in a single session consisting of 14 experimental runs divided in 2 baseline recordings (eyes open and closed) and 3 runs per experimental condition. The experimental conditions consisted of the motor execution or imagination of left and right hand movements or both hands and feet movements.\\
The signal was acquired with a sampling rate of 160Hz and no pre-processing was performed on the data.

\subsection{MI-OpenBCI}
The \textit{MI-OpenBCI} dataset \cite{peterson2020feasibility} \cite{peterson2022motor} has been collected and provided to the research community very recently (year 2020).\\
The authors performed a feasibility study on the use of consumer-grade MI EEG-based BCI system. Therefore, they employed the OpenViBE software platform\footnote{\url{http://openvibe.inria.fr/}.}, the OpenBCI Cyton sensing board with Daisy Module\footnote{\url{https://docs.openbci.com/GettingStarted/Boards/DaisyGS/}.} and used Electrocap System II with 19 electrodes to acquire the EEG signal wirelessly. Of these electrodes, 16 were effectively used (F\{z,3,4,7,8\}, C\{z,3,4\}, T\{3,4,5,6\}, P\{z,3,4\}). Reference and ground electrodes were placed on the left and right ear lobes.\\
Notice that electromyographic signals were also acquired, but details are not provided here, having that these physiological data are not the focus of the present work.\\
12 healthy right-handed subjects with no prior experience with BCIs performed a cue-based MI of the dominant hand grasping or resting during a single experimental session. 4 runs of 20 trials were executed.\\
The EEG signal was acquired with sampling rate of 125Hz and filtered with a 3rd order Butterworth bandpass-filter (0.5-45Hz). 

\section{Artificial intelligence approaches on the investigated datasets}\label{sec:ai}
Having provided a brief description of each dataset under scrutiny, it is possible to proceed with the analysis of the AI approaches applied on them during their lifetime. \\
The analyses will start from the first publication presenting results on the investigated dataset and then proceed by considering some relevant works that allow further discussions on the influence that AI has on MI EEG-based BCIs. 

\subsection{On the BCI Competition IV dataset 2a}\label{sec:bciiv2a}
The winners of the \textit{BCI Competition IV dataset 2a} presented their approach in a dedicated publication \cite{ang2012filter}. \textit{Ang et al.} wanted to enhance the performances of the Common Spatial Pattern (CSP) algorithm, and thus proposed the use of Filter Bank CSP (FBCSP) \cite{ang2008filter}. FBCSP consists of 4 phases, i.e., the band-pass filtering, the spatial filtering, the feature selection based on mutual information, and the classification phase. Concerning this last step, the authors chose to apply the Naïve Bayesian Parzen Window classifier, having obtained good results from a previous competition \cite{ang2008filter}. 
Moreover, having that the dataset presented 4 conditions and that the algorithm was initially developed for binary classification, they propose multi-class extensions of FBCSP by using the divide and conquer, pairwise and one versus rest approaches. This last strategy was chosen as the one to present to the competition having that it obtained better or similar average Cohen's kappa value (0.57) with less computational cost, datum that is particularly important when thinking of the real-time nature that BCI responses should have.

In 2015, \textit{Nicolas-Alonso et al. }\cite{nicolas2015adaptive} thought about this issue and proposed an adaptive semi-supervised classification that could also face the non-stationarity of the EEG signals. Therefore, they firstly reduced non-stationarity by using exponentially weighted moving average before performing the classification, which is done by using the newly developed method based on spectral regression kernel discriminant analysis. In fact, the authors train the model with labelled data and then introduce a self-training algorithm using new unlabelled data to sequentially update the model. 
Notice that the feature extraction is done through CSP.\\
The joint semi-supervised learning and adaptive processing provided better results with a significant computational efficiency compared to previous literature works considering the 4-class problem (0.70 Cohen's kappa coefficient and 77\% accuracy).\\
The authors further improved the performances (0.74 maximum average kappa value) by using the stacked regularised linear discriminant analysis for classification and thus integrating temporal, spectral, and spatial information \cite{nicolas2015adaptiveIEEE}.

An adaptive learning strategy has been also proposed by \textit{Raza et al.} \cite{raza2016adaptive}, in 2016, who exploited the covariate shift detection based on exponentially weighted moving average to monitor the EEG-based BCIs. The authors obtained an average accuracy of 76.70\% employing the upper bound version of their methodology.

In 2017, \textit{Jafarifarmand et al.} \cite{jafarifarmand2017new} proposed a new framework that could allow the improvement of multi-class BCIs. The process consisted of a feature extraction step using artifact rejected CSP and a self-regulated
adaptive resonance theory based neuro-fuzzy classifier that is able to model the non-stationarity of the EEG signal as well as uncertainties in the data. The authors obtained better results in respect to the competition winners with an average kappa value of 0.63.

In 2019, \textit{Olivas-Padilla} \cite{olivas2019classification} do not only test their strategy on \textit{BCI Competition IV dataset 2a} but also acquire a proprietary dataset using OpenBCI with a similar experimental paradigm. The feature extraction was again provided by a novel modification of the CSP algorithm, i.e., the discriminative FBCSP, to model the spatial and spectral characteristics of the EEG signal. A modular network composed of 4 Convolutional Neural Networks (CNNs) specialised in the binary classification of combinations of the 4 experimental conditions, was used to provide a final classification. Notice that Bayesian optimisation was employed for hyperparameter selection. The authors' model achieved 80.03\% accuracy and a kappa score of 0.61.

In 2019, deep learning methods were also investigated by \textit{Majidov \& Whangbo} \cite{majidov2019efficient} who highlighted the necessity of these models to use a good number of data. Therefore, they proposed a pipeline consisting of different combinations of the following steps: (i) a data augmentation step based on sliced moving windows, (ii) the extraction of the power spectrum density from 3 EEG rhythms and the application of FBCSP, (iii) information theoretical feature extraction, (iv) tangent space mapping feature extraction and, (v) wrapper-based feature-selection with particle swarm optimization. The classification was performed with a 1D CNN, avoiding deeper networks that could overfit and obtaining an average accuracy of 87.94\%.

In the same year (2019), 83\% accuracy and 0.80 Cohen's kappa value were instead achieved by \textit{Zhang et al.} \cite{zhang2019novel}. The authors proposed the use of one-versus-rest FBCSP to extract characterising features for each of the 4 classes and a deep architecture based on the CNN and Long Short Term Memory (LSTM) models. Notice that they performed the classification by training the model on the data merged from all the subjects and then evaluating the data of each subject separately, obtaining a subject-invariant strategy.

A year later (2020), \textit{Luo et al.} \cite{luo2020motor} proposed a novel ensemble support vector learning based approach to combine event related synchronisation/desynchronisation features typical of the MI domain. The authors employed class discrepancy-guided sub-band filter-based common spatial pattern to extract more separable and stable features before the classification step. The proposed approach achieved an average kappa value of 0.60.

Finally, in 2021, a CNN model built on the inception-time network called \textit{EEG-inception} \cite{zhang2021eeg} and, in 2022, a transfer learning-based CNN and LSTM hybrid deep learning model \cite{khademi2022transfer} have been  proposed. \textit{EEG-inception} is directly fed with the raw EEG data augmented with noise addition, avoiding complex pre-processing steps and decreasing the possibility of overfitting. 88.39\% average accuracy was obtained on the 4 classes.\\
Similarly, \textit{Khademi et al.} \cite{khademi2022transfer} firstly applied a data augmentation by cropping and secondly provided a time-frequency characterization of the EEG signal through continuous wavelet transform. Afterwards, the authors exploited ResNet-50 and Inception-v3, 2 pre-trained CNNs, to provide a transfer learning strategy supporting their CNN/LSTM model. The mean kappa values obtained by using the provided approach were 0.86 and 0.88 for the ResNet-50 and Inception-v3 models, respectively. The accuracy values were around 90\% and 92\%.

Table~\ref{tab:bci_civ_2a} summarises the time-line of the presented AI strategies and highlights (bold) the best achieved results.

\begin{table}
	\caption{\textit{BCI Competition IV dataset 2a} time-line on AI strategies summary. Best results are highlighted (bold).}
	\label{tab:bci_civ_2a}
	\resizebox{\textwidth}{!}{%
	\begin{tabular}{ccp{7.5cm}cc}
		\hline
		Paper & Year & Strategy & Cohen's kappa & Accuracy \\
		\hline
		\cite{ang2012filter} & 2012 & FBCSP $+$ Naïve Bayesian Parzen Window classifier & 0.57 & NA \\
		\cite{nicolas2015adaptive} & 2015 & CSP $+$ adaptive semi-supervised classification & 0.70 & 77.00\% \\
		\cite{nicolas2015adaptiveIEEE} & 2015 & Added on \cite{nicolas2015adaptive} stacked regularised linear discriminant analysis & 0.74 & NA \\
		\cite{raza2016adaptive} & 2016 & Adaptive learning with covariate shift detection & NA & 76.70\% \\
		\cite{jafarifarmand2017new} & 2017 & Artifact rejected CSP $+$ self-regulated neuro-fuzzy framework & 0.63 & NA \\
		\cite{olivas2019classification} & 2019 & FBCSP $+$ 4 CNNs $+$ Bayesian optimisation & 0.61 & 80.03\% \\
		\cite{majidov2019efficient} & 2019 & Data augmentation $+$ feature engineering $+$ 1D CNN & NA & 87.94\% \\ 
		\cite{zhang2019novel} & 2019 & One-vs-rest FBCSP $+$ CNN and LSTM (subject-invariant) & \textbf{0.80} & 83.00\% \\
		\cite{luo2020motor} & 2020 & CSP variation $+$ ensemble support vector learning & 0.60 & NA \\
		\cite{zhang2021eeg} & 2021 & \textit{EEG-inception} & NA & 88.39\% \\
		\cite{khademi2022transfer} & 2022 & Data augmentation $+$ Inception-v3 & NA & \textbf{92.00\%} \\
		\hline
	\end{tabular}
}
\end{table}

\subsection{On the BCI Competition IV dataset 2b}
A good number of works are in common between the two BCI Competition IV datasets under analysis. In fact, the winners using the \textit{BCI Competition IV dataset 2b} were again \textit{Ang et al.} \cite{ang2012filter}, who used the same approach described for \textit{BCI Competition IV dataset 2a} at the beginning of Section~\ref{sec:bciiv2a}. They obtained an average accuracy around the 60\%.\\
Similarly, the work of \textit{Raza et al.} \cite{raza2016adaptive} appears again in the performed search. The average accuracy obtained on dataset 2b was of 73.33\% using the upper bound version of the proposed strategy.

In 2019, \textit{Zhu et al.} \cite{zhu2019separated} proposed an end-to-end deep learning framework based on the transfer of knowledge obtained from previously analysed subjects with the main aim of removing the training phase from MI BCIs. CSP was used to extract features in the temporal domain and the authors introduced a separated-channel CNN to provide the correct characterisation of the multi-channel EEG signal. Besides testing the proposed strategy on the scrutinised dataset, they performed analyses on a proprietary recorded set of data, considering left/right hand MI. Notice that a single subject data was used as the test set, while the remaining subject data were exploited for model training. A comparison with widely used traditional machine learning techniques was also performed considering the K-nearest neighbour, logistic regression, linear discriminant analysis and Support Vector Machine (SVM) classifiers. Therefore, close performances were detected between the novel approach without transfer learning and the traditional models, especially considering the information transfer rate on the BCI Competition IV dataset 2b. Instead, better average results were provided when the transfer learning was applied. The proposed approach achieved 0.83 Information Transfer Rate (ITR), while the best result for the traditional techniques was obtained by the SVM with an ITR value of 0.02. The accuracy values were around 64\% and 50\% for the two strategies, respectively.

Instead, in the same year, \textit{Malan \& Sharma} \cite{malan2019feature} proposed a paradigm shift by considering a better definition of the feature vector to enhance a SVM classifier performances. Therefore, after extracting features by considering frequency-related and statistical measures, the authors performed feature selection by proposing a regularised neighbourhood component analysis. In fact, the authors wanted to reduce the feature vector by selecting the features providing the maximum accuracy or minimum generalisation error.\\
The final results show that the proposed algorithm was able to provide better average performances in terms of accuracy (80.70\%), kappa coefficient (0.62), precision (0.85), recall (0.79), specificity (0.83) and F1-score (0.81) by using a lower number of features (6 on 42) when compared with feature selection techniques based on ReliefF, principal component analysis and genetic algorithm.

An year later (2020), the ensemble support vector learning strategy of \textit{Luo et al.} \cite{luo2020motor} described in Section~\ref{sec:bciiv2a} provided 0.71 average max kappa value on the analysed dataset.\\ 
Even \textit{Zhang et al.} \cite{zhang2021eeg} (2021) tested their \textit{EEG-inception} model on both the BCI Competition IV datasets and obtained the 88.60\% average accuracy on dataset 2b.

Finally, \textit{Malan \& Sharma} \cite{malan2022motor} presented a new work in 2022, focusing again on the optimisation of the feature vector to improve a SVM classifier performances. Therefore, they proposed a novel methodology consisting of (i) dual tree complex wavelet transform based filter bank, (ii) CSP for spatial feature extraction from the previously obtained EEG sub-bands, and again (iii) a regularised neighbourhood component analysis. Comparing their approach with other CSP variations, they obtained better results in terms of accuracy (84\%) and kappa coefficient (0.68).

Table~\ref{tab:bci_civ_2b} summarises the time-line of the presented AI strategies and highlights (bold) the best achieved results.

\begin{table}
	\caption{\textit{BCI Competition IV dataset 2b} time-line on AI strategies summary. Best results are highlighted (bold).}
	\label{tab:bci_civ_2b}
	\resizebox{\textwidth}{!}{%
	\begin{tabular}{ccp{7.5cm}cc}
		\hline
		Paper & Year & Strategy & Cohen's kappa & Accuracy \\
		\hline
		\cite{ang2012filter} & 2012 & FBCSP $+$ Naïve Bayesian Parzen Window classifier & NA & 60.00\% \\
		\cite{raza2016adaptive} & 2016 & Adaptive learning with covariate shift detection & NA & 73.33\% \\
		\cite{zhu2019separated} & 2019 & CSP $+$ knowledge transfer $+$ CNN & NA & 64.00\% \\
		\cite{malan2019feature} & 2019 & Feature selection through regularised neighbourhood component analysis & 0.62 & 80.70\% \\
		\cite{luo2020motor} & 2020 & CSP variation $+$ ensemble support vector learning & \textbf{0.71} & NA \\
		\cite{zhang2021eeg} & 2021 & \textit{EEG-inception} & NA & \textbf{88.60\%} \\
		\cite{malan2022motor} & 2022 & Feature vector optimisation $+$ SVM & 0.68 & 84.00\% \\
		\hline
	\end{tabular}
}
\end{table}

\subsection{On the EEG Motor Movement/Imagery Dataset}
The \textit{EEG Motor Movement/Imagery Dataset} has not been used as much as the previous 2 datasets, but provides a good starting point to perform analyses on a larger pool of data considering both executed and imagined movements.\\
The first work by \textit{Sleight et al.} (2009) \cite{sleight2009classification} reporting results on this dataset performed a particularly interesting analysis having that the authors' aim was not to discriminate different MI tasks, but the imagined from the real movements. Therefore, they firstly considered feature extraction based on independent component analysis on different frequency bands and a channel selection, to provide a well characterised feature vector as input to a SVM model. After performing different analyses, the best average accuracy (69\%) was obtained by considering a subject-based approach on data normalized per frequency band and avoiding the use of independent component analysis.

Almost 5 years later, in 2013, \textit{Park et al.} \cite{park2013augmented} proposed the augmented complex CSP to deal with non-circular EEG signals. The SVM model (with a Gaussian kernel) was employed to classify data from a synthetic dataset and the one under scrutiny, considering left/right hand MI. Among the 109 subjects, 56 performed over the 64\% accuracy.

In 2017, a LSTM model was proposed \cite{zhang2017intent} to obtain intent recognition considering all the experimental MI conditions of 10 subjects. A hyper-parameter selection through orthogonal array application was also performed obtaining a final average accuracy of 95.53\%.\\
Another deep learning approach has been chosen in 2018 by \textit{Dose et al.} \cite{dose2018end}, who considered the combination of different MI tasks for signal classification, i.e., left/right hand, open eyes with left/right hand, feet with open eyes and left/right hand conditions. The authors designed a CNN architecture that could be complemented with a transfer learning strategy. The average accuracy values obtained on the different task combinations without (with) the transfer learning approach were 80.38\% (86.49\%), 69.82\% (79.25\%) and 58.58\% (68.51\%), respectively. A general performance increase was detected when applying the transfer learning strategy.

An ensemble learning approach for MI data classification on 10 subjects was instead chosen by \textit{Zhang et al.} \cite{zhang2018converting} in 2018. Firstly, the authors exploited a recurrent neural network and a CNN model to learn features and then an autoencoder for feature adaption. Afterwards, they applied eXtreme Gradient Boosting for intent recognition, obtaining an average accuracy of 95.53\%.

The prediction of different MI conditions (left/right hand, plus rest, plus feet) was also provided by \textit{Wang et al.} \cite{wang2020accurate} (2020), who exploited a model based on a pre-trained CNN, \textit{EEG-Net}, obtaining the following accuracy values: 82.43\%, 75.07\% and 65.07\%.

Finally, in 2021, \textit{Varsehi \& Firoozabadi} \cite{varsehi2021eeg} focused on a channel selection method, taking as an assumption the causal interaction between channels. Therefore, the authors proposed a novel channel selection algorithm based on Granger causality analysis and after different processing steps, classified the resulting data by using the linear/quadratic discriminant analysis, kernel Fisher discriminant, SVM, multi-layer perceptron, learning vector quantisation, neural network, k-nearest neighbour and decision tree classifiers. The best results obtained by the proposed strategy and using 8 channels only were of 93.03\% accuracy, 92.93\% sensitivity, and 93.12\% specificity.

Table~\ref{tab:physionet} summarises the time-line of the presented AI strategies and highlights (bold) the best achieved results.

\begin{table}
	\caption{\textit{EEG Motor Movement/Imagery Dataset} time-line on AI strategies summary. Best results are highlighted (bold).}
	\label{tab:physionet}
	\resizebox{\textwidth}{!}{%
	\begin{tabular}{ccp{7.5cm}cc}
		\hline
		Paper & Year & Strategy & Cohen's kappa & Accuracy \\
		\hline
		\cite{sleight2009classification} & 2009 & Imagined vs real movements $+$ independent component analysis $+$ SVM  & NA & 69.00\% \\
		\cite{park2013augmented} & 2013 & Augmented complex CSP $+$ SVM (56 subjects) & NA & 64.00\%\\
		\cite{zhang2017intent} & 2017 & LSTM (all MI conditions) $+$ hyperparameter selection through orthogonal array & NA & \textbf{95.53\%} \\
		\cite{dose2018end} & 2018 & CNN (left/right hand MI) $+$ transfer learning & NA & 86.49\% \\
		\cite{zhang2018converting} & 2018 & Recurrent neural network $+$ CNN for feature extraction $+$ eXtreme Gradient Boosting (all MI conditions) & NA & \textbf{95.53\%} \\
		\cite{wang2020accurate} & 2020 & \textit{EEG-Net} (left/right hand MI) & NA & 82.43\%\\
		\cite{varsehi2021eeg} & 2021 & channel selection based on Granger causality & NA & 93.03\% \\
		\hline
	\end{tabular}
}
\end{table}

\subsection{On the MI-OpenBCI dataset}
Being very recent (2020), the \textit{MI-OpenBCI} dataset has not yet been deeply analysed by the research community.\\
The dataset authors, \textit{Peterson et al.},  \cite{peterson2020feasibility} proposed data denoising and feature extraction based on CSP. The best results were achieved by the penalised time-frequency band CSP with an average accuracy of 83.30\%. Notice that generalised sparse discriminant analysis was used for both feature selection and classification considering both online and offline scenarios.\\
Therefore, the proposed methodology is set as a benchmark for future works.\\
Notice that the works citing \textit{Peterson et al.} paper have exploited it to make a point regarding the necessity of moving to low-cost and consumer-grade technologies to provide more ecological and easy-to-use BCIs.\\
For example, \textit{Koo et al.} \cite{koo2021demonstration} cite it when considering that locked-in patients do not have communication tools based on BCI paradigms that are easy to maintain, low cost and comfortable. Moreover, \textit{Peterson et al.} work is taken as an example considering the use of open-source and low cost platforms \cite{butsiy2021comprehensive} \cite{zambrana2022validation} like the OpenBCI one.

\section{Discussion}\label{sec:discussion}
Considering the general overview given on the AI techniques applied to the analysed datasets, observations can be made towards the scope of the present paper.
\begin{enumerate}
	\item Initially, the BCI Competition IV datasets were considered as the sole test beds for novel MI detection techniques. Afterwards, researchers began to use them as terms of comparison for their proprietary data. The \textit{EEG Motor Movement/Imagery Dataset} seemed to immediately take the role of a benchmark.
	\item The CSP and its variations have been widely used and remain between the most applied techniques for feature extraction, meant to provide a correct data characterisation as input to AI models.
	\item From an initial application of statistical measures or SVM classifiers, the AI techniques evolved following this pattern: (i) traditional machine learning approaches, (ii) introduction of transfer or adaptive learning, (iii) focus on the feature vector and channel selection, (iv) deep learning architectures, especially based on CNN and LSTM, or ensemble techniques, and (v) exploitation of pre-trained networks.
	\item The necessity of introducing data augmentation techniques has also appeared when the researches conducted with deep learning approaches became more mature.
	\item The check on performances has shifted from the use of the sole kappa coefficient to an extreme tendency of accessing the system behaviour through accuracy only.
	\item New technologies and demands from the general public have asked a paradigm shift related to the use of low-cost and easy-to-use devices, which should enable a rapid response from BCI systems and thus provide lightweight computation techniques to allow efficient and reliable feedback to the users as part of their requirements.
	\item The acquisition of the \textit{MI-OpenBCI} dataset has set a good standard to provide systems in line with the requirements reported at the previous point.
\end{enumerate}
Therefore, the evolution of AI techniques is clearly detectable and followed the discoveries presented by the AI community. However, this evolution demanded changes in how the experimental paradigms and devices are made. Especially, the need for more data to feed to the models has become particularly pressing as well as the possibility of providing signals with higher signal-to-noise ratio.\\
Interestingly, a return to simpler and quicker traditional machine learning models in combination with feature engineering strategies, may prove effective for new real-time and low-cost BCI systems. In fact, instead of answering the demands from a computational perspective, the user needs will probably become the main focus when developing new technologies. 

\section{Conclusion}\label{sec:conclusion}
In this paper, a brief over all survey on the AI techniques applied to specific MI EEG-based BCI datasets has been provided. \\
Starting from the development of the BCI2000 platform and the opening of the BCI Competitions, great interest has been given by the field research community towards the generation of new approaches enabling the correct response of such systems.\\
In the presented overview, the evolution of AI techniques and their influence on MI EEG-based BCIs seemed to have been twofold: the AI approaches have provided better discrimination of experimental conditions and the devices used to collect data and control applications have changed according to the needs of the general public.\\
However, many issues remain to be answered, starting from the necessity of finding reliable metrics for feedback procedure evaluations not only based on the sole accuracy but also accessing the usability, comfort, reliability and reproducibility of the new consumer-grade technologies.

\bibliographystyle{splncs04}
\bibliography{mybib}


\end{document}